\documentclass[nofootinbib,showpacs,superscriptaddress,11pt]{revtex4-1}

\usepackage{graphicx,natbib}
\usepackage{amsmath,bm,mathrsfs}

\begin{document}

\title{On the Intensity Distribution Function\break
of Blazed Reflective Diffraction Gratings}

\author{R.\ Casini}
\affiliation{High Altitude Observatory,
National Center for Atmospheric Research,\break
P.\ O.\ Box 3000, Boulder, CO 80307-3000, U.S.A.}

\author{P.\ G.\ Nelson}

\begin{abstract}
We derive from first principles the expression for the angular/wavelength 
distribution of the intensity diffracted by a blazed reflective grating, 
according to a scalar theory of diffraction. We considered the 
most common case of a groove profile with rectangular apex.
Our derivation correctly identifies the geometric parameters of a blazed 
reflective grating that determine its diffraction efficiency, and fixes 
an incorrect 
but commonly adopted expression in the literature. We compare the
predictions of this scalar theory with those resulting from a rigorous 
vector treatment of diffraction from one-dimensional blazed reflective 
gratings.
\end{abstract}


\maketitle

\section{Introduction} \label{sec:intro}

Diffraction gratings are widely used tools for astronomical 
applications, in both ground-based and space-borne telescope
facilities. A proper implementation of gratings for spectroscopic 
observations requires a full understanding of their properties.
However, apart from a few results following directly from the 
grating equation (spectral dispersion, free spectral range, blaze 
wavelength), the derivation of other important characteristics of 
gratings is less straightforward.
This is certainly the case for the determination of the
angular/wavelength distribution of the diffracted intensity, which 
is needed for the estimation of the grating efficiency under specific 
illumination conditions, and for given diffraction orders and 
directions. 
A reliable estimate of the grating efficiency is critical in the 
design of spectroscopic tools to be employed for observations that 
are particularly demanding on the photon flux reaching the detector. 
A typical example is that of high-sensitivity spectro-polarimetry, 
which is a fundamental diagnostic tool for the inference of magnetic 
fields in astrophysical plasmas.

A correct determination of the grating intensity distribution must 
take into account the polarization properties of gratings, and can only 
be attained within a full (vector) electro-magnetic theory of diffraction. 
However, the simpler scalar theory is often very useful, being capable 
of providing a good approximation to the average (i.e., unpolarized) 
efficiency, which can be adopted for reliable flux-budget estimations in 
spectrographic instruments. This is particularly true for small blaze 
angles, and for small ratios of the wavelength to the grating period
\cite{LP97}.
Echelle gratings with steeper blaze angles (e.g., of the
R2 type, which is commonly adopted in high spectral resolution 
instrument setups for the remote sensing of astrophysical plasmas), 
working in relatively low orders, typically display strong polarization 
features and anomalies. However, special reflective coating techniques 
such as ``shadow casting'' \cite{Ke66} have been demonstrated to 
effectively reduce these anomalies to such a level that the scalar theory 
of diffraction becomes again useful also for the modeling of the efficiency
of this type of gratings.

In the case of transmission gratings, the expression for 
the intensity distribution of the diffracted radiation is well-known 
\cite{Gr05,Sa06,Se10}, and it depends in a fundamental way on the grating 
period, $d$, and the width of the transmitting aperture, $b$. 
Unfortunately, the commonly used extension of that formula to blazed 
reflective gratings appears to be marred by a confused identification of the 
width $b$ \cite{Gr05}. This misuse of the intensity distribution 
function typically results in grossly overestimated or underestimated 
efficiencies for echelle gratings, ultimately affecting the reliability 
of the design of spectrographic instruments. \cite{Sc00} provides 
the correct identification of the $b$ parameter for blazed reflective
gratings under different illumination conditions, although without formal
derivation.

\begin{figure}[t]
\centering
\includegraphics[width=\hsize]{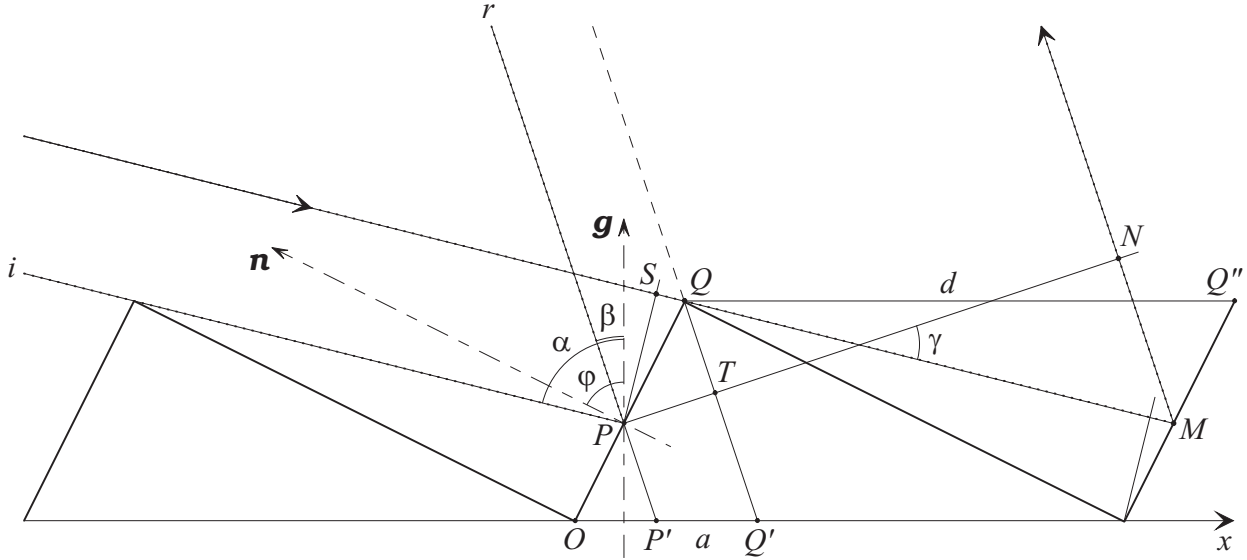}
\caption{\label{fig:grating}
Geometric construction for the derivation of the grating efficiency
formula. The grating shown in the figure has ruling step $d$ and 
blaze angle $\varphi$, with the two facets forming a right angle 
(in $Q$). $\alpha$ is the angle of incidence of the incoming light
($i$) measured from the normal $\bm{g}$ of the grating, while $\beta$ 
gives the direction of the diffracted light ($r$). $\bm{n}$ is the 
normal to the illuminated facet, so that $\varphi=\widehat{\bm{gn}}$.}
\end{figure}

In Section~\ref{sec:theory}, we derive from first principles the scalar 
intensity distribution function of a blazed reflective grating, for the
most common case of a groove profile with rectangular apex.
Comparing our result with the commonly adopted expression for this
intensity distribution, we attain a clear identification of the
geometric parameters involved. This derivation of the intensity
distribution function for a blazed reflective grating is validated by 
the results
of numerical modeling based on a vector theory of diffraction from
one-dimensional diffraction gratings, which are presented in 
Section~\ref{sec:discussion}, as well as from laboratory measurements 
(H.~Lin, private communication).

\section{Theory} \label{sec:theory}

Figure~\ref{fig:grating} shows the geometric construction for a 
reflective diffraction grating with a blaze angle $\varphi$ and 
ruling period $d$.
For simplicity, we assume the most common case of a grating profile with 
rectangular apex. Conventionally, we indicate with $\alpha$ the
incidence angle of the incoming radiation ($i$) with respect to the grating
normal $\bm{g}$, and with $\beta$ the corresponding angle for the
diffracted radiation ($r$).
In the following derivation we assume the condition $\alpha\ge\varphi$, 
which ensures that the secondary facet is never illuminated (and 
therefore does not contribute to the diffracted energy). The case 
$\alpha<\varphi$ will be discussed briefly at the end, in the context 
of \textit{shadowing}.

Following \cite{Gr05}, the distribution of the diffracted intensity by
a grating, according to the scalar theory of diffraction, can be
obtained from the Fourier transform of the grating transmission (or
aperture) function, $G(x)$, where $x$ is the coordinate along the 
grating plane (see Fig.~\ref{fig:grating}). 
This transmission function is different from zero only in correspondence 
of the \textit{openings} of the grating, 
i.e., the regions of the $x$-axis from where the diffracted radiation appears
to emanate. These openings are the analog of the 
transmitting apertures in a transmission grating. For a blazed reflective
grating, such as the one pictured in Fig.~\ref{fig:grating}, we see that
these openings correspond to the segment $P'Q'$ and periodic replicas of 
it along the grating length. The width and location of these segments
are evidently functions of the angles $\alpha$ and $\beta$.
In particular, the width is given by
\begin{equation} \label{eq:P'Q'}
a\equiv P'Q'=PQ\,\frac{\cos(\beta-\varphi)}{\cos\beta}\;,
\end{equation}
whereas the position of the median points of the openings is located at
$\bar{x}=[x(P')+x(Q')]/2$ and periodic replicas of it along the grating 
length, where
\begin{eqnarray} \label{eq:eq2}
\bar{x} = x(P')+\frac{a}{2} 
&=&d\sin\varphi\tan(\alpha-\varphi)
	(\cos\varphi+\sin\varphi\tan\beta) 
  +\frac{d}{2}\,\frac{\cos\alpha}{\cos\beta}\,
	\frac{\cos(\beta-\varphi)}{\cos(\alpha-\varphi)} \nonumber \\
&=&
  \frac{d}{2}\,\frac{\cos(\alpha-2\varphi)}{\cos\beta}\,
	\frac{\cos(\beta-\varphi)}{\cos(\alpha-\varphi)}\;.
\end{eqnarray}

In order to properly calculate the Fourier transform of the transmission
function $G(x)$, we need to determine the phase dependence on $x$ of the
diffracted wave across the width $P'Q'$. In order to do so, we 
must calculate the path difference between the incident and the
diffracted wave after the plane $PS$, at which the incident (plane) wave 
is in phase. The segment $PQ$ illuminated by the incident light is given by
\begin{eqnarray} \label{eq:PQ}
PQ
&=&d\cos\varphi-d\sin\varphi \tan(\alpha-\varphi) \nonumber \\
&=& d\,\frac{\cos\alpha}{\cos(\alpha-\varphi)}\;.
\end{eqnarray}
Therefore, from Fig.~\ref{fig:grating}, the path differences for the 
diffracted beam are
\begin{subequations}
\begin{eqnarray}
SQ
&=&PQ\sin(\alpha-\varphi)\;, \\
TQ
&=&PQ\sin(\varphi-\beta)\;.
\end{eqnarray}
\end{subequations}
The overall phase difference between the points $T$ and $P$ in the
diffracted wave for the wavelength $\lambda$ is thus
\begin{eqnarray} \label{eq:phase_TP}
\theta(\varphi)
&=&\frac{2\pi}{\lambda}\,(SQ-TQ) \nonumber \\
&=&\frac{2\pi}{\lambda}\,PQ\,[\sin(\alpha-\varphi)
	+\sin(\beta-\varphi)]\;,
\end{eqnarray}
We note that for
\begin{equation} \label{eq:eq6}
\alpha-\varphi=\varphi-\beta\;,
	\quad\Longrightarrow\quad
	\alpha+\beta=2\varphi\;,
\end{equation}
the phase $\theta(\varphi)$ vanishes identically. This is the ordinary
condition of reflection on the grating facets ({\it blaze condition\/}).

For values of $x$ intermediate between $P'$ and $Q'$, the phase
difference between the diffracted ray through $x$ and the 
one through $\bar{x}$ scales with the factor
\begin{equation}
\rho_x\equiv\frac{x-\bar{x}}{ a}\;,\qquad
	x(P')\le x\le x(Q')\;,
\end{equation}
which therefore must be introduced into Eq.~(\ref{eq:phase_TP}), as a 
multiplicative scaling factor of $PQ$, in order to calculate the 
dephasing of the diffracted beam introduced by the blaze. 

The derivation given above allows the computation of the phase dependence
of an arbitrary diffracted ray through the segment $P'Q'$ for the 
groove that was conventionally set as the origin. 
Each successive groove introduces an offset of the phase retardance 
that is constant throughout the grating (i.e., independent of $x$), 
and which is determined by the additional travel of the plane wave 
over the path $SQ+QM+MN$. The segment $SQ$ has already been determined,
whereas
\begin{equation} \label{eq:QM}
QM=d\,\frac{\sin\varphi}{\cos(\alpha-\varphi)}\;.
\end{equation}
Since $\gamma=90^\circ-(\alpha-\beta)$, we also have
\begin{equation}
TQ+MN=QM\cos(\alpha-\beta)\;,
\end{equation}
where $TQ$ has also been determined earlier. The phase retardance
introduced by each groove is therefore
\begin{eqnarray}
\Theta
&=&\frac{2\pi}{\lambda}\,(SQ+QM+MN) \nonumber \\
&=&\frac{2\pi}{\lambda}\,[SQ-TQ+QM+(TQ+MN)] \nonumber \\
&=&\frac{2\pi}{\lambda}
	\Bigl\{PQ\,[\sin(\alpha-\varphi)+\sin(\beta-\varphi)] 
	+QM\,[1+\cos(\alpha-\beta)]\Bigr\}\;,
\end{eqnarray}
with $PQ$ and $QM$ given by Eqs.~(\ref{eq:PQ}) and (\ref{eq:QM}), 
respectively. With these
substitutions, after some tedious algebra, we find simply
\begin{equation} \label{eq:grating_eq}
\Theta=\frac{2\pi}{\lambda}\,d\,(\sin\alpha+\sin\beta)\;.
\end{equation}
The last expression can be cast into the usual form of the grating 
equation (e.g., \cite{Gr05}), when we observe that the different orders 
$n$ of diffraction by the grating must correspond to phase conditions of 
constructive interference, i.e., $\Theta=2\pi n$, with $n$ an integer.
%

We now consider the explicit expression of the transmission function,
$G(x)$, for a blazed grating. This is given by (cf.~\cite{Gr05}, 
Eq.~(3.3))
\begin{equation}
G(x)=a(x)\ast{\rm III}_d(x-\bar{x})\,U_L(x)\;,
\end{equation}
where $a(x)$ is the ``window'' function associated with the segment
$P'Q'$ of width $a$, ${\rm III}_d(x)=\sum_n \delta(x-nd)$ is the 
sampling function (Dirac's \emph{comb}) of the grating, and $U_L(x)$ is the 
box function of unit height that limits the total length, $L$, of 
the grating. For a blazed grating, $a(x)$ is not purely real, since 
it carries the additional dephasing due to the blaze with respect to 
the case of a flat grating. For the interval $x-\bar{x}$ this 
dephasing is evidently given by (see Eqs.~(\ref{eq:eq2}) and 
(\ref{eq:eq6}))
\begin{equation} \label{eq:dephasing}
\Delta\theta(x)\equiv\biggl[\frac{\theta(\varphi)}{ a}
	-\frac{\Theta}{ d}\biggr](x-\bar{x})\;.
\end{equation}
If $U_a(x)$ is the unit box of width $a$, then
\begin{equation}
a(x)=U_a(x)\,\hbox{e}^{{\rm i}\,\Delta\theta(x)}\;.
\end{equation}
The Fourier transform\footnote{For ease of comparison, we adopt the
same sign convention of \cite{Gr05} for the argument of the exponential 
phase factor in the Fourier transform integral.} of $G(x)$ is given by
\begin{eqnarray} \label{eq:Gtransform}
\tilde{G}&\equiv&\mathscr{F}\bigl\{G(x)\bigr\} \nonumber \\
&=&\mathscr{F}\bigl\{a(x)\bigr\}
	\Bigl[
	\mathscr{F}\bigl\{{\rm III}_d(x-\bar{x})\bigr\}\ast
	\mathscr{F}\bigl\{U_L(x)\bigr\}
	\Bigr]\;,\kern .5cm
\end{eqnarray}
and must be evaluated at $\sigma=(\sin\alpha+\sin\beta)/\lambda
\equiv \Theta/(2\pi d)$. Equation~(\ref{eq:dephasing}) can then be 
rewritten,
\begin{equation}
\Delta\theta(x)=\biggl[\frac{\theta(\varphi)}{ a}
	-2\pi\sigma\biggr](x-\bar{x})\;.
\end{equation}

Using fundamental properties of the Fourier transform, for the various 
contributions to Eq.~(\ref{eq:Gtransform}), we find:
\begin{subequations}
\begin{eqnarray} \label{eq:fourier_1}
\kern -1cm
\mathscr{F}\bigl\{a(x)\bigr\}(\sigma) 
&=&\hbox{e}^{{\rm i}\,\Delta\theta(0)}\,
	\mathscr{F}\biggl\{U_a(x)
	\exp\biggl({\rm i}\,\biggl[\frac{\theta(\varphi)}{ a}
	-2\pi\sigma\biggr]x\biggr)\biggr\}(\sigma) \nonumber \\
&=&\hbox{e}^{{\rm i}\,\Delta\theta(0)}\,
	\mathscr{F}\bigl\{U_a(x)\bigr\}
	\biggl(\frac{\theta(\varphi)}{ 2\pi a}\biggr) \nonumber \\
&=&\hbox{e}^{{\rm i}\,\Delta\theta(0)}\,
	{\rm sinc}\biggl(\frac{\theta(\varphi)}{ 2}\biggr)\;;
\end{eqnarray}
\begin{eqnarray} \label{eq:fourier_2}
\mathscr{F}\bigl\{{\rm III}_d(x-\bar{x})\bigr\}(\sigma)
&=&\hbox{e}^{{\rm i}\,2\pi\sigma\bar{x}}\,
	\mathscr{F}\bigl\{{\rm III}_d(x)\bigr\}(\sigma) \nonumber \\
&=&\hbox{e}^{{\rm i}\,2\pi\sigma\bar{x}}\,
{\rm III}_{1/d}(\sigma)\;;
\end{eqnarray}
\begin{equation} \label{eq:fourier_3}
\mathscr{F}\bigl\{U_L(x)\bigr\}(\sigma)={\rm sinc}(\pi L\sigma)\;.
\end{equation}
\end{subequations}
Ultimately, we are only interested in the intensity distribution
function of the diffracted field, which is proportional to 
$|\tilde G|^2$, so the phase factors in Eqs.~(\ref{eq:fourier_1}) and
(\ref{eq:fourier_2}) can simply be dropped. 
We note that Eq.~(\ref{eq:fourier_2}) determines the free spectral 
range of the grating as a dispersing element, whereas 
Eq.~(\ref{eq:fourier_3}) determines its \textit{finesse} (or resolution).
The ``envelope'' of the diffracted light distribution is instead
exclusively determined by Eq.~(\ref{eq:fourier_1}) (cf.~\cite{Gr05},
Eqs.~(3.4) and (3.8)), and this is the quantity we are interested in for
the present study. 

If we recall Eqs.~(\ref{eq:PQ}) and (\ref{eq:phase_TP}), we find from
Eq.~(\ref{eq:fourier_1})
\begin{eqnarray} \label{eq:formula}
I(\beta)
&=&{\rm sinc}^2\biggl(
	\frac{\pi d}{\lambda}\,\frac{\cos\alpha}{\cos(\alpha-\varphi)}\,
	[\sin(\alpha-\varphi)+\sin(\beta-\varphi)]
	\biggr) \nonumber \\
&=&{\rm sinc}^2\biggl(
	n\pi\,\frac{\cos\alpha}{\cos(\alpha-\varphi)}\,
	\frac{\sin(\alpha-\varphi)+\sin(\beta-\varphi)}{
	\sin\alpha+\sin\beta}
	\biggr) \nonumber \\
&=&{\rm sinc}^2\biggl(
	n\pi\,\frac{\cos\alpha}{\cos(\alpha-\varphi)}
	\biggl[\cos\varphi-\sin\varphi\cot\frac{\alpha+\beta}{2}\biggr]
	\biggr)\;, \nonumber \\
\end{eqnarray}
where in the second equivalence we used the grating equation
(cf.~Eq.~(\ref{eq:grating_eq}), and the discussion following that
equation).
If we recall the discussion following Eq.~(\ref{eq:phase_TP}),
Eq.~(\ref{eq:formula}) shows that, according to the scalar theory 
of gratings, the peak of the efficiency is reached at the blaze 
condition, $\alpha+\beta=2\varphi$. For a given grating configuration
and diffraction order $n$, the efficiency peak then occurs at the 
so-called \textit{blaze wavelength}
\begin{equation} \label{eq:blaze}
\lambda_{\rm b}\equiv\frac{d}{n}\,[\sin\alpha+\sin(2\varphi-\alpha)]\;.
\end{equation}
We also note that, for $\lambda=\lambda_{\rm b}$, the Littrow 
condition, $\alpha=\beta$, also corresponds to the configuration of 
normal incidence on the grating facets, $\alpha=\varphi$, in which case
\begin{equation} \label{eq:Littrow}
\lambda_n\equiv\lambda_{\rm b}^{\rm Litt}=\frac{2d}{n}\,\sin\varphi\;.
\end{equation}

Comparison of Eq.~(\ref{eq:formula}) with Eq.~(3.8) of \cite{Gr05}, 
shows that it must be, for $\alpha\ge\varphi$,
\begin{equation} \label{eq:b/d}
\frac{b}{ d}=\frac{\cos\alpha}{\cos(\alpha-\varphi)}\;,
\end{equation}
where $b$ is the effective width of the openings that must
be adopted according to \cite{Gr05} in order to reproduce the correct
envelope of the diffracted energy $I(\beta)$. From this we conclude 
that $b=PQ$, according to our derivation, i.e., 
\textit{the effective width of the openings in a blazed reflective
grating corresponds to the width of the illuminated portion of the 
grating's facet} (see Eq.~(\ref{eq:PQ})).
Equation~(\ref{eq:b/d}) is in agreement with Eq.~(13.4.10) of
\cite{Sc00}. 

In the condition of normal incidence on the grating
facets, $\alpha=\varphi$, we have simply $b/d=\cos\varphi$. \cite{Gr05} 
reports instead $b/d=\cos^2\varphi$, suggesting that the author identifies 
$b$ with the \textit{normal projection} of the illuminated portion of
the facet onto the grating length 
(i.e., with the quantity $a=P'Q'$ of Fig.~\ref{fig:grating}). 
In the following, we will indicate
this as ``Gray's \textit{ansatz}.'' Of course, the difference
between these two conflicting identifications of the parameter $b$ may 
lie below the limit of experimental detection for small blaze
angles. Instead, in the case of echelle gratings, the difference in the 
estimated grating efficiencies provided by the two alternate formulations 
can be significant.\footnote{For the case of $\varphi=20^\circ$ considered 
by \cite{Gr05}, the two definitions of $b/d$ imply a difference of 
$\sim 6\%$ in the argument of the $\rm sinc^2$ function. For an
echelle grating with $\varphi=60^\circ$, the arguments would differ by
a factor 2 instead.\label{fn:shallow}} In the next Section, we present 
comparisons with results from a rigorous vector model of grating 
diffraction, which support the validity of Eq.~(\ref{eq:b/d}).

Interestingly, the same problem has also been treated by \cite{Br59} 
through a geometric argument similar to the one presented here. 
The author arrives at an expression for the ratio $b/d$ that is formally 
identical to Eq.~(\ref{eq:b/d}), but where $\beta$ appears instead of 
$\alpha$. While this would still lead to the correct estimation of the 
$b/d$ ratio at the Littrow condition, in the general case it determines 
an unphysical behavior of the diffracted efficiency towards the 
wavelength corresponding to the ``passing off'' of a diffraction order,
which ultimately violates energy conservation. In fact, it can be 
demonstrated that, for $\beta\to\pi/2$, Eq.~(\ref{eq:formula}) tends 
identically to unity, if $\alpha$ and $\beta$ are exchanged in that 
expression.

From the geometric construct of Fig.~\ref{fig:grating} we can also 
determine the grating magnification (also known as anamorphic 
magnification), which is defined as $r=PS/PN$. From the above derivation, 
we have $PS=PQ\cos(\alpha-\varphi)$, and 
$PN=PT+TN=PQ\cos(\varphi-\beta)+QM\sin(\alpha-\beta)$. After some simple 
algebra, we find the usual result
\begin{equation} \label{eq:magnification}
r=\frac{\cos\alpha}{\cos\beta}\;.
\end{equation}

\begin{figure}[t!]
\includegraphics[width=\hsize]{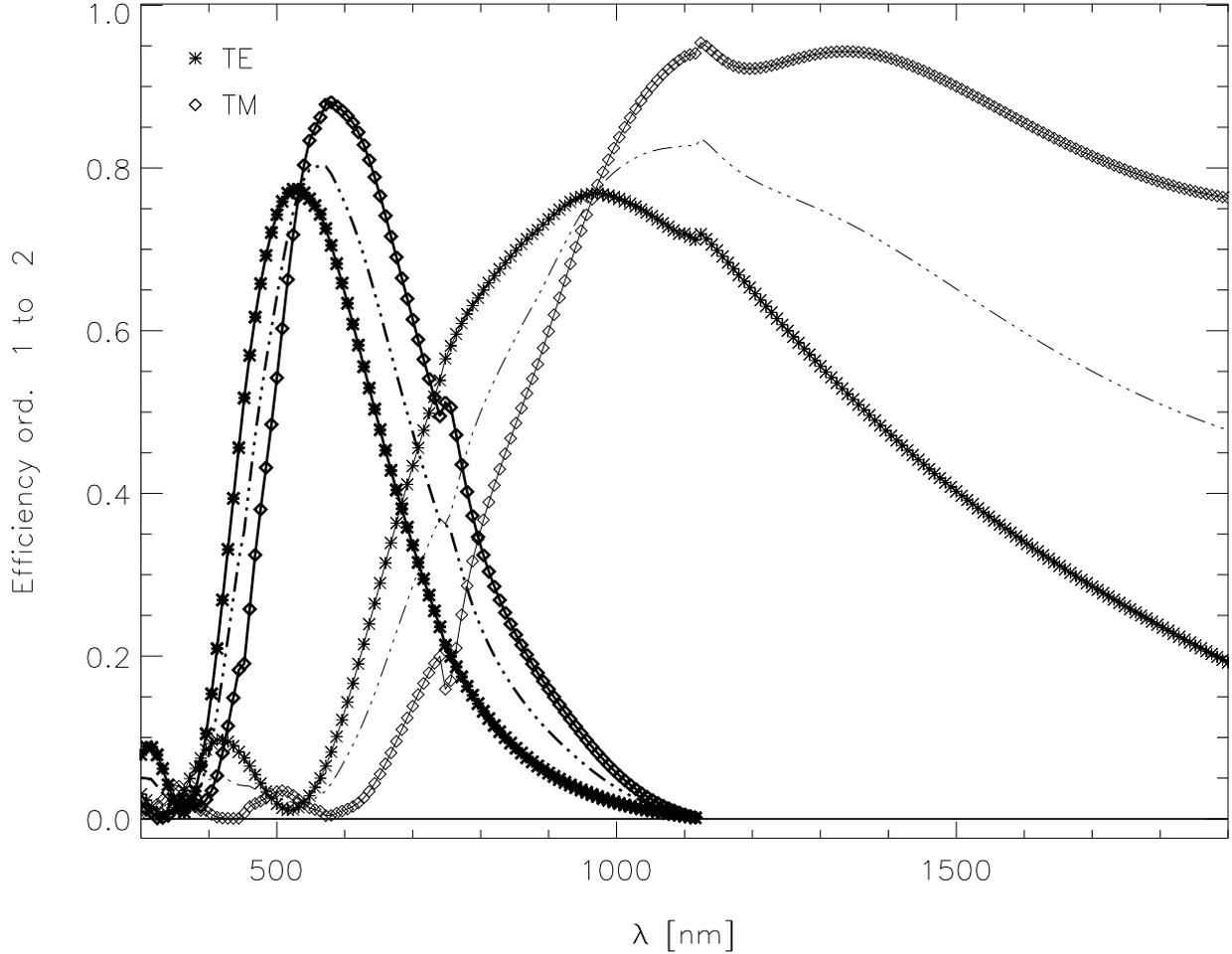}
\caption{Efficiency curves for the orders 1 (thin curves) and 2 
(thick curves) for an Al grating with 600~lines/mm and $\varphi=20^\circ$, 
used in the configuration of normal incidence on the grating facets, 
$\alpha=\varphi$. The theoretical TE ($p$) and TM ($s$) polarizations 
are shown, as well as the unpolarized case (corresponding to the 
average of the two polarizations; dash-dotted curves).}
\label{fig:vector_eff.ja}
\end{figure}

\begin{figure}[t!]
\includegraphics[width=\hsize]{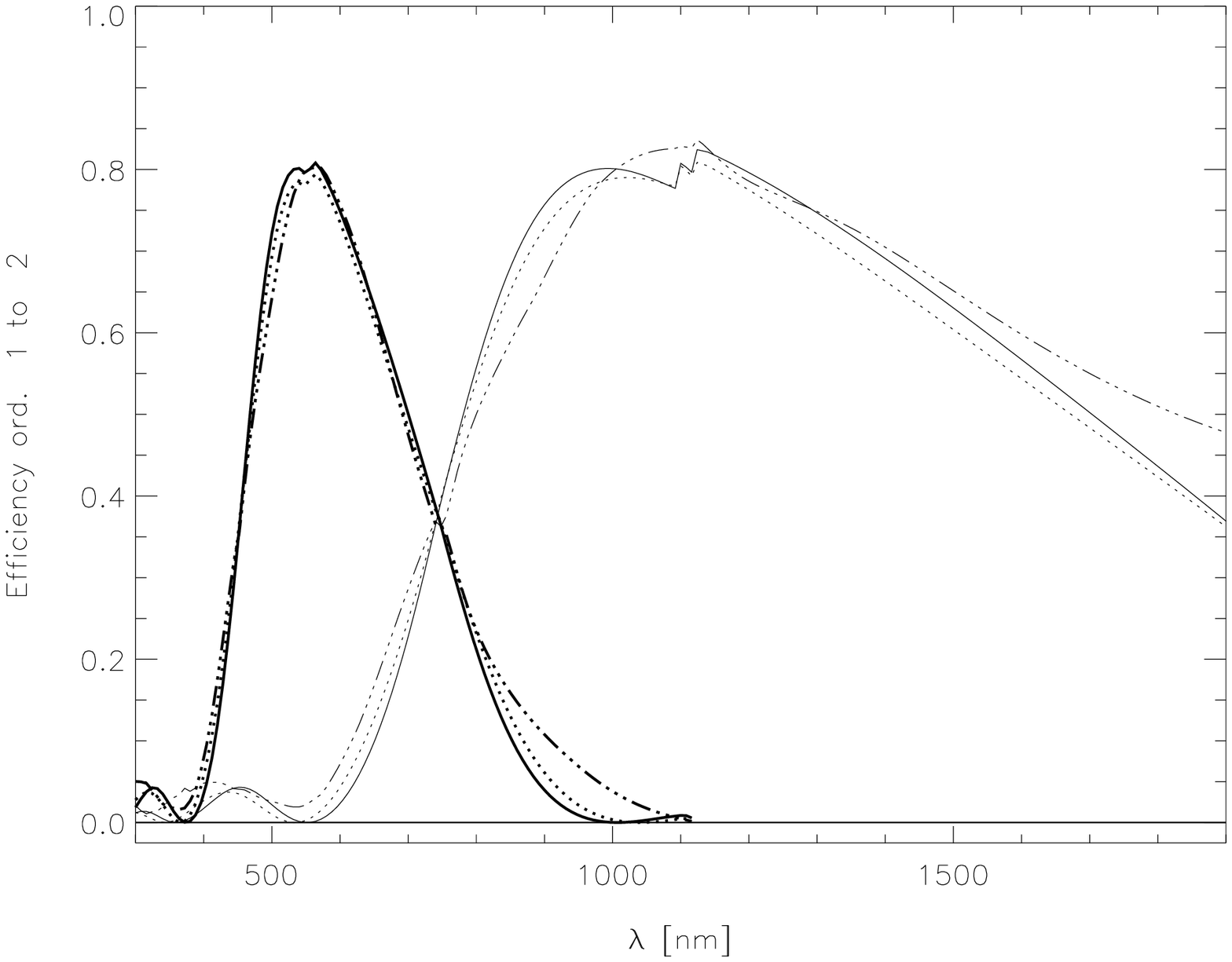}
\caption{Same as Fig.~\ref{fig:vector_eff.ja}, but using the scalar theory
derived in this paper (continuous curves). An overall grating loss of 
16.5\% was introduced via a scaling factor, in order to match the efficiency 
peaks of the continuous curves to those of the unpolarized efficiency
curves of Fig.~\ref{fig:vector_eff.ja} (dash-dotted curves in this plot). 
The dotted curves correspond to the grating efficiency determined by 
Gray's \textit{ansatz} for the expression of the ratio $b/d$ (see 
Eq.~(\ref{eq:b/d})). We note how the differences introduced by Gray's 
\textit{ansatz} are very small in this case, because of the small blaze 
angle.}
\label{fig:comparison.ja}
\end{figure}

Finally, the grating model given in Fig.~\ref{fig:grating} can also
be used to determine the efficiency loss due to \textit{shadowing}, 
which sets in when the illumination of the grating is such that 
$\alpha<\beta$. We first consider the case $\alpha<\varphi$, and
apply the principle of reversibility in order to exchange the role 
of $\alpha$ and $\beta$ in Fig.~\ref{fig:grating}. We then can conclude
that only a fraction $P'Q'/d\equiv a/d$ of the incoming light is 
diffracted into the outgoing direction. From Eqs.~(\ref{eq:P'Q'}) 
and (\ref{eq:PQ}), and taking into
account the reversed roles of $\alpha$ and $\beta$, the reduction 
factor of the grating efficiency due to shadowing is therefore
\begin{eqnarray} \label{eq:shadowing}
s\equiv a/d
&=&\frac{\cos\beta}{\cos\alpha}\,
	\frac{\cos(\alpha-\varphi)}{\cos(\beta-\varphi)} \nonumber \\
&=&\frac{1}{ r}\,\frac{\cos(\alpha-\varphi)}{\cos(\beta-\varphi)}\;.
\end{eqnarray}
where in the last line we used the definition of the grating
magnification, Eq.~(\ref{eq:magnification}). It is easy to demonstrate
that Eq.~(\ref{eq:shadowing}) applies also to the case 
$\alpha\ge\varphi$. 
To see this, we must picture the geometric construct of 
Fig.~\ref{fig:grating} for the case $\beta>\alpha$, and 
observe that the unblocked portion of the diffracted beam 
appears to originate from a sub-region of the grating
facet of width $P''Q$, such that (cf.~Eq.~(\ref{eq:PQ}))
\begin{equation}
P''Q=d\,\frac{\cos\beta}{\cos(\beta-\varphi)}
<PQ=d\,\frac{\cos\alpha}{\cos(\alpha-\varphi)}\;.
\end{equation}
In this case, the reduction factor of the grating efficiency
due to shadowing is $s\equiv P''Q/PQ<1$, and the last equation shows 
that the expression of $s$ coincides with the RHS of 
Eq.~(\ref{eq:shadowing}) also for $\alpha\ge\varphi$.

Equation~(\ref{eq:shadowing}) shows that shadowing reduces the 
\emph{peak efficiency} of a grating order
by a factor $1/r=\cos\beta/\cos(2\varphi-\beta)$, because of the blaze 
condition $\alpha-\varphi=\varphi-\beta$. This result is in agreement with 
the treatment of shadowing given in \cite{Gr05}, as seen from Eq.~(3.11) 
in that reference. However, that expression strictly holds only at the 
blaze condition, unlike Eq.~(\ref{eq:shadowing}) in this paper, which is 
instead general. 

In conclusion, the expression of the grating efficiency, 
Eq.~(\ref{eq:formula}), must in general be multiplied by a scaling
factor $k=\min(s,1)$, with $s$ given by Eq.~(\ref{eq:shadowing}). This
reduction factor sets in under the shadowing condition $\alpha<\beta$, 
and applies identically in the two cases $\alpha\ge\varphi$ and 
$\alpha<\varphi$.

We then can rewrite the expression of the grating efficiency for
the wavelength $\lambda$ or the diffraction order $n$ as
\begin{eqnarray} \label{eq:formula.final}
I(\beta)&=&
k\;{\rm sinc}^2\biggl(
	\frac{\pi d}{\lambda}\,\rho\,
	[\sin(\alpha-\varphi)+\sin(\beta-\varphi)]
	\biggr) \nonumber \\
&=&
k\;{\rm sinc}^2\biggl(
	n\pi\,\rho
	\biggl[\cos\varphi-\sin\varphi\cot\frac{\alpha+\beta}{2}\biggr]
	\biggr)\;,\kern .5cm
\end{eqnarray}
where $\rho\equiv b/d$ is given by Eq.~(\ref{eq:b/d}) for 
$\alpha\ge\varphi$, while it remains at the maximum possible value 
$\cos\varphi$ for $\alpha<\varphi$.


\begin{figure}[t!]
\includegraphics[width=\hsize]{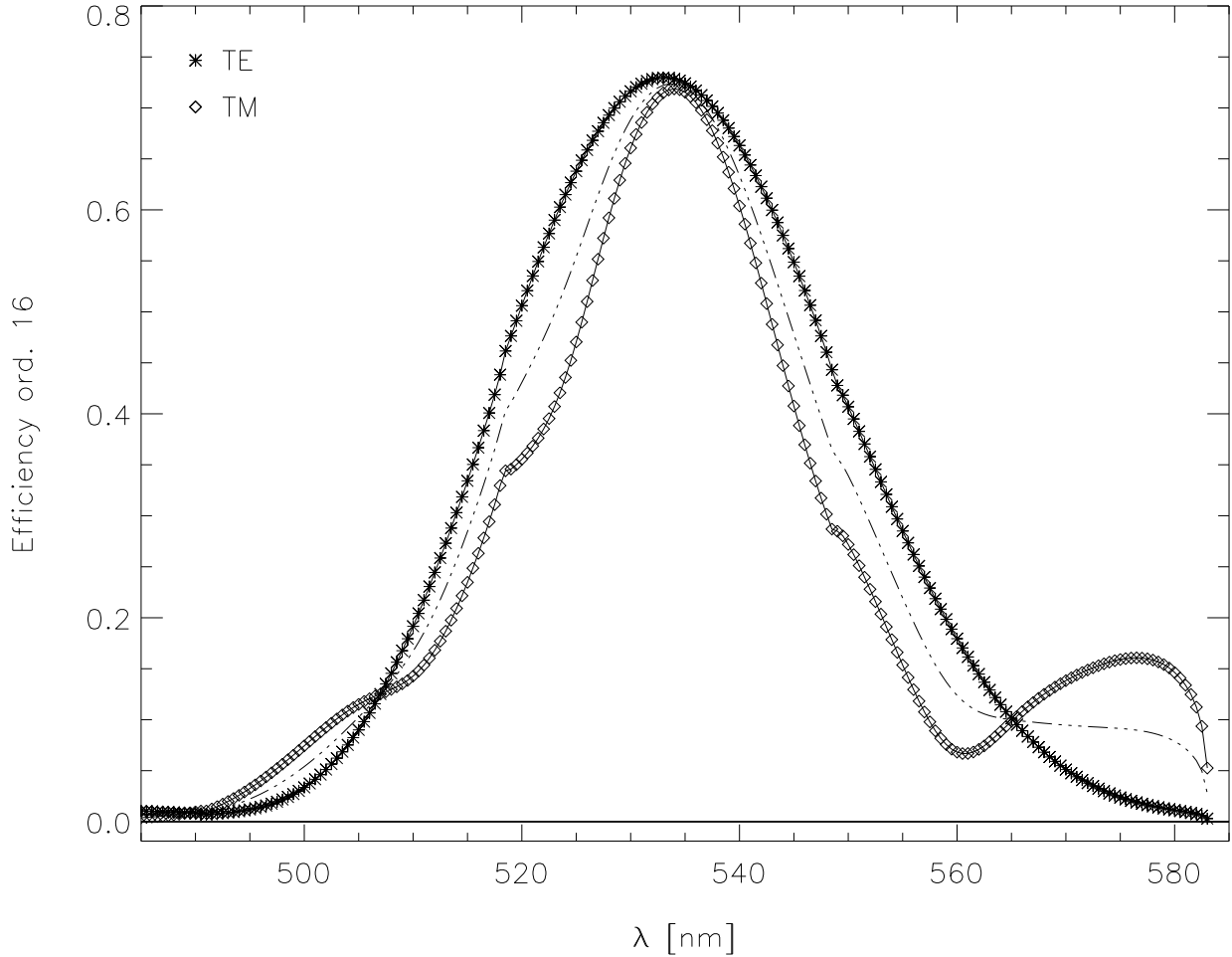}
\caption{Same as Fig.~\ref{fig:vector_eff.ja}, but for the order 16 
of an Ag grating with 200~lines/mm and $\varphi=60^\circ$,
and under the illumination condition $\alpha=\varphi$.}
\label{fig:vector_eff}
\end{figure}

\begin{figure}[t!]
\includegraphics[width=\hsize]{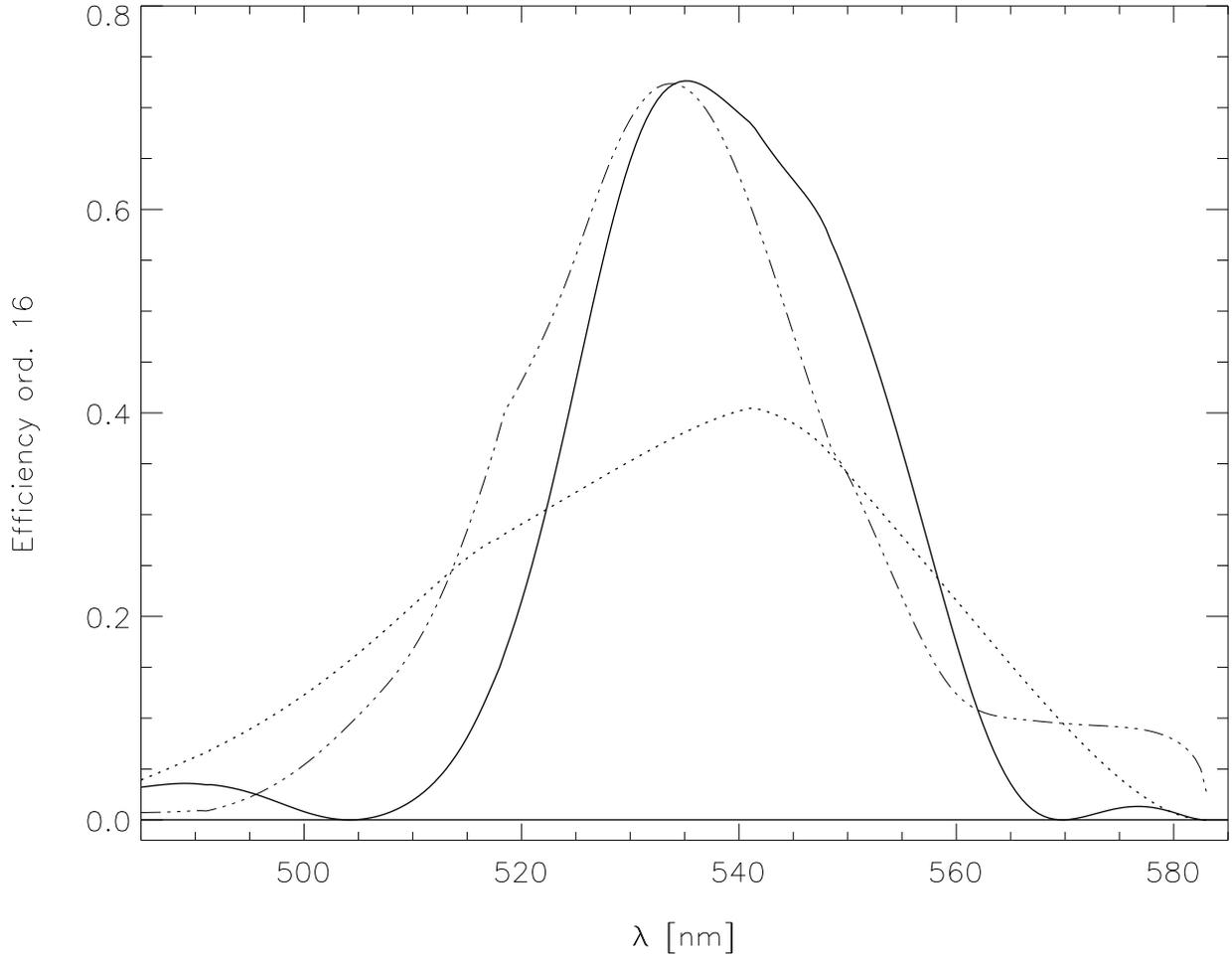}
\caption{Same as Fig.~\ref{fig:vector_eff}, but using the scalar
theory as for the plots of Fig.~\ref{fig:comparison.ja} (continuous
curve). An overall 
grating loss of 20\% due to the reflective layer was considered for 
this calculation. In this case,
the scalar theory is able to reproduce the unpolarized efficiency 
of the grating (dash-dotted curve), at least around the peak, if 
Eq.~(\ref{eq:b/d}) is adopted for $b/d$. In contrast, the scalar 
theory is unable to reproduce both efficiency and position of 
the peak, as well as the FWHM of the
efficiency profile, if Gray's \textit{ansatz} is used instead
(dotted curve).}
\label{fig:comparison}
\end{figure}

\section{Discussion and Conclusions} \label{sec:discussion}

In this section we will consider some examples of grating efficiency
calculations, in order to test the ability of the scalar theory of 
gratings presented above to reproduce results predicted by a rigorous 
(vector) treatment of light diffraction by one-dimensional gratings.
 
The formulation presented above reproduces rather well the average
efficiency of blazed reflective gratings as derived from a full 
treatement of the electro-magnetic theory of diffraction, at the 
condition that energy 
conservation across the various orders and efficiency losses due to 
shadowing are taken into account, and that polarization effects
introduced by the grating are not predominant. 
A good reproduction of the peak efficiency may need an additional scaling 
factor to account for the wavelength dependence of the reflectivity of 
the coating. Since this reflectivity does not enter explicitly the
scalar theory of gratings, it must be introduced \textit{ad hoc} as a
normalization parameter.

Figure \ref{fig:vector_eff.ja} shows the efficiency curves for the 
TE (or $p$) and TM (or $s$) polarizations of the diffracted field in the
orders $n=1,2$ of an Al grating with 600~lines/mm and 
$\varphi=20^\circ$, used in the configuration of normal incidence on the 
grating facets (i.e., $\alpha=\varphi$). We recall that such
configuration corresponds to a Littrow mount ($\alpha=\beta$) when
$\lambda=\lambda_n$ (see Eq.~(\ref{eq:Littrow})). The plots of 
Fig.~\ref{fig:vector_eff.ja} were calculated with a code based on the 
C-method for grating analysis as described in \cite{Li99}. For comparison, 
the continuous curves in Fig.~\ref{fig:comparison.ja} represent the 
scalar efficiencies calculated through Eq.~(\ref{eq:formula.final}),
taking into account energy conservation, the energy 
loss due to shadowing as described by Eq.~(\ref{eq:shadowing}), and 
adding an overall grating loss of 16.5\%. We note that in this 
case the scalar theory is able to reproduce adequately the position of 
the efficiency peak, as well as the full width at half maximum (FWHM) 
and overall trend of the unpolarized efficiency profile. The dotted 
curves in Fig.~\ref{fig:comparison.ja} show the predicted efficiency 
using Gray's \textit{ansatz} for the $b/d$ ratio (see discussion after 
Eq.~(\ref{eq:b/d})). For this grating -- which is analogous to the one
considered by \cite{Gr05} in his Figs.~3.11 and 3.12 -- the differences 
between the two alternate definitions of $b/d$ are very small, as 
expected because of the small blaze angle of the grating (see 
note~\ref{fn:shallow}).

%

Figures~\ref{fig:vector_eff} and \ref{fig:comparison} provide another
test of the performance of the scalar theory of gratings. These new
calculations are for a Ag grating with 200~lines/mm and a blaze angle 
$\varphi=60^\circ$. Also in this case, the grating is used in a
configuration of normal incidence on the grating facets, which
corresponds to a Littrow mount at the blaze wavelength 
$\lambda_n\approx 541.3$\,nm for the case $n=16$ shown in the figure.

Comparing Figs.~\ref{fig:vector_eff} and \ref{fig:comparison} we see 
that for echelle gratings the agreement between the scalar and vector 
theories of diffraction is significantly worse than in the case 
illustrated by Figs.~\ref{fig:vector_eff.ja} and \ref{fig:comparison.ja}. 
Nonetheless, the scalar theory is still capable of reproducing the 
efficiency curve in a neighborhood of the peak, as well as the FWHM 
of the efficiency profile, which is an important quantity for a correct 
estimation of the bandwidth of the diffraction orders. In contrast, use 
of Gray's \textit{ansatz} for echelle gratings gives results that are 
completely at variance with those of the vector theory. In particular, 
because of the much larger (by $\sim 60$\%) FWHM of the efficiency 
profile determined by Gray's \textit{ansatz}, the overlap between 
distinct orders at any given wavelength is also much larger than
in reality, so it becomes impossible to reproduce the peak efficiency 
simply because of energy conservation -- i.e., the diffracted energy at 
any given wavelength gets distributed into too many orders. The position 
of the efficiency peak also misses to reproduce the results of the 
vector theory in this case, remaining practically located at 
$\lambda_n$.


\begin{acknowledgments}
We are grateful to H.~Lin (IfA, U.\ of Hawaii) for several discussions 
about this problem at an early stage of this work, and to A.~de Wijn 
(HAO, NCAR) for a careful reading of the present manuscript and helpful 
comments. We express our thanks to the anonymous reviewer of the paper 
for many insightful comments and suggestions.
The National Center for Atmospheric Research is sponsored by
the National Science Foundation.
\end{acknowledgments}

\end{document}